\documentclass[fleqn,usenatbib]{mnras}

\usepackage{newtxtext,newtxmath}
\usepackage[T1]{fontenc}
\usepackage{ae,aecompl}

\usepackage{graphicx}	% Including figure files
\usepackage{amsmath}	% Advanced maths commands
\usepackage{amssymb}	% Extra maths symbols

\title[MHD Wire Turbulence]{Hydrodynamic and Magnetohydrodynamic Simulations of Wire Turbulence}

\author[E. Fogerty et al.]{Erica Fogerty$^{1,2}$\thanks{E-mail:efogerty@lanl.gov}, Baowei Liu$^{3,4,5}$, Adam Frank$^{3}$, Jonathan Carroll-Nellenbeck$^{3,4,5}$, \newauthor Sergey Lebedev$^{6}$ \\\\
$^{1}$ Center for Theoretical Astrophysics, Los Alamos National Laboratory, Los Alamos, New Mexico\\
$^{2}$ Computer, Computational, and Statistical Sciences Division, Los Alamos National Laboratory, Los Alamos, New Mexico\\
$^{3}$ Department of Physics \& Astronomy, University of Rochester, Rochester, New York\\
$^{4}$ Center for Integrated Research Computing, University of Rochester, Rochester, New York\\
$^{5}$ Laboratory for Laser Energetics, University of Rochester, Rochester, New York\\
$^{6}$ Department of Physics, Imperial College London, Kensington, London SW7 2AZ}

\pubyear{2019}

\begin{document}
\label{firstpage}
\pagerange{\pageref{firstpage}--\pageref{lastpage}}
\maketitle

\begin{abstract}
We report on simulations of laboratory experiments in which magnetized supersonic flows are driven through a wire mesh. The goal of the study was to investigate the ability of such a configuration to generate supersonic, MHD turbulence. We first report on the morphological structures that develop in both magnetized and non-magnetized cases. We then analyze the flow using a variety of statistical measures, including power spectra and probability distribution functions of the density.  Using these results we estimate the sonic mach number in the flows downstream of the wire mesh.  We find the initially hypersonic ($M_s=20$) planar shock through the wire mesh does lead to downstream turbulent conditions.  However, in both magnetized and non-magnetized cases, the resultant turbulence was marginally supersonic to transonic ($M_s \sim 1)$, and highly anisotropic in structure.
\end{abstract}


%%%%%%%%%%%%%%%%%%%%%%%%%%%%%%%%%%%%%%%%%%%%%%%%%%

%%%%%%%%%%%%%%%%% BODY OF PAPER %%%%%%%%%%%%%%%%%%

\section{Introduction}

Supersonic, magnetohydrodynamic (MHD) turbulence occurs in many astrophysical settings, from star formation in the interstellar medium \citep{maclow2004, federrath2012, kritsuk2017, offner2018}, to supernova engines \citep{couch2015, fryer2017, radice2018} and remnants \citep{balsara2001,inoue2009, roy2009}, to the solar wind \citep{Alexandrova2008, bruno2013, Usmanov2014}. As such, theoretical and simulation-based studies of MHD turbulence have been a robust endeavor within the astrophysical community, articulating important properties of MHD turbulence with regard to turbulent power spectra, decay rates and observational characteristics \citep{elmegreen2004, padoan2004, kritsuk2011b, Federrath2015, kritsuk2017}. A problem for simulation-based studies of MHD turbulence, however, has been the limited range of Reynolds numbers (both hydrodynamic and magnetic) achievable with even modern numerical codes. Typically these values are many orders of magnitude smaller than what would be expected for real astrophysical flows (c.f. Elmegreen and Scalo 2004). %Values of $Re \sim 10^3$ and $Re_M \sim 100^2$ are typical for simulation based studies (REF).  These are many orders of magnitude smaller than what is expected for real astrophysical flows.

Over the last two decades, high energy density laboratory astrophysics (HEDLA) studies have opened new paths to the study of astrophysical phenomena \citep{2006RvMP...78..755R}.  The stability of supernova blast waves \citep{2002ApJ...564..896D}, magnetized jets \citep{2002ApJ...564..113L, 2007PhPl...14e6501C, 2009PhPl...16d1005B, 2015ApJ...815...96S} and shock-clump interactions \citep{2016ApJ...823..148H} have all been usefully explored in laboratory settings using high energy density platforms, such as high intensity lasers and pulse power machines.  Laser-based experiments have recently shown the potential for HEDLA studies to explore issues of MHD turbulence. For example, \cite{Meinecke2015} used colliding plasma jets to study the development of Kolmogorov-like
turbulence, and showed that the magnetic field in the flows was amplified by turbulent motions. Earlier work by \cite{2014NatPh..10..520M} showed that MHD turbulence could be achieved after a laser driven plasma flow generated from a carbon rod had been passed through a grid. Experiments to generate MHD turbulence could also be performed using magnetized plasma flows generated with pulsed power drivers \citep{Lebedev2014, Bott-Suzuki2015, Burdiak2017,Lebedev2019}

In this paper, we report on simulations that also used a grid to generate turbulence from an initially laminar flow.  Using the adaptive mesh refinement (AMR), MHD code AstroBEAR, we tracked the evolution of the flow to explore the conditions under which turbulence could be generated. The work and setup in this paper is intended to provide guidance for future studies of laboratory supersonic turbulence. The paper is organized as follows. The numerical methods and simulations are described in Section \ref{section:methods}, Results are presented in Section \ref{section: results}, and a discussion of the findings are given in Section \ref{section: discussion}. 

\section{Methods and Simulation Parameters}\label{section:methods}

To explore the possible generation of supersonic turbulence in optically thin plasma in a laboratory setting, we conducted a set of high resolution, 3D simulations of a hypersonic wind colliding with a wire mesh.  Our simulations were carried out using the AstroBEAR\footnote{See: https://astrobear.pas.rochester.edu/trac/ for a listing of current capabilities.} code \citep{cunningham2009, carroll13}, a state-of-the-art, multiphysics AMR platform for solving the equations of hydrodynamics/MHD in the Eulerian frame.  Two simulations were performed of a hypersonic wind being passed through a wire mesh -- one in  which the wind was hydrodynamic, and the other in which the wind carried a magnetic field. For the MHD case, the simulations solved the ideal MHD equations,

\begin{equation}
    \frac{\partial \rho}{\partial t} + \boldsymbol{\nabla} \cdot (\rho \boldsymbol{v}) = 0
    \label{eq:Eu1}
\end{equation}

\begin{equation}
    \frac{\partial (\rho \boldsymbol{v})}{\partial t} + \boldsymbol{\nabla} \cdot \left ( \rho \boldsymbol{v} \boldsymbol{v} +P\boldsymbol{I}-\boldsymbol{B}\boldsymbol{B}\right )=0
    \label{eq:Eu2}
\end{equation}
\begin{equation}
    \frac{\partial E}{\partial t} + \boldsymbol{\nabla} \cdot [(E + P) \boldsymbol{v}-(\boldsymbol{v}\cdot\boldsymbol{B})\boldsymbol{B}] = 0
    \label{eq:Eu3}    
\end{equation}

\begin{equation}
    \frac{\partial \boldsymbol{B}}{\partial t} + \boldsymbol{\nabla} \cdot ( \boldsymbol{v}\boldsymbol{B}-\boldsymbol{v}\cdot\boldsymbol{B})\boldsymbol{B}) = 0
    \label{eq:Eu4}    
\end{equation}

\noindent where $\rho$ is the mass density, $\boldsymbol{v}$ is the velocity, $P$ is the thermal pressure, $\mathbf{B}$ is the magnetic field, and $E$ is the total energy, given by $E = \frac{p}{\rho(\gamma - 1)} +  \frac{1}{2}\rho v^{2}+\frac{1}{2}B^{2}$ (note, the hydro simulation solved an identical set of equations, but with $\mathbf{B}=0$). To close the system of equations, an ideal gas equation of state was used with an adiabatic index very close to one ($\gamma=1.001$). Setting $\gamma\approx 1$ effectively treats the gas as isothermal. Gases typically used in laboratory experiments of wire turbulence cool efficiently through radiative loses. Thus, treating the gas isothermally allowed us to approximate radiative cooling without explicitly including radiative processes in the calculations. 

In each of the simulations, a hypersonic wind was injected into a computational domain filled with a single gas at two different densities. The different densities were used to represent a dense wire mesh embedded within a sparse ambient gas (gas that made up the wires was $10^4$ times denser than the surrounding ambient medium). Since the wires simply represented over dense regions in the flow (i.e. did not carry an electrical current), wire material could be ablated over the course of the simulation through interaction with the incident hypersonic flow. 

\begin{figure*}
\centering
\includegraphics[width=\textwidth]{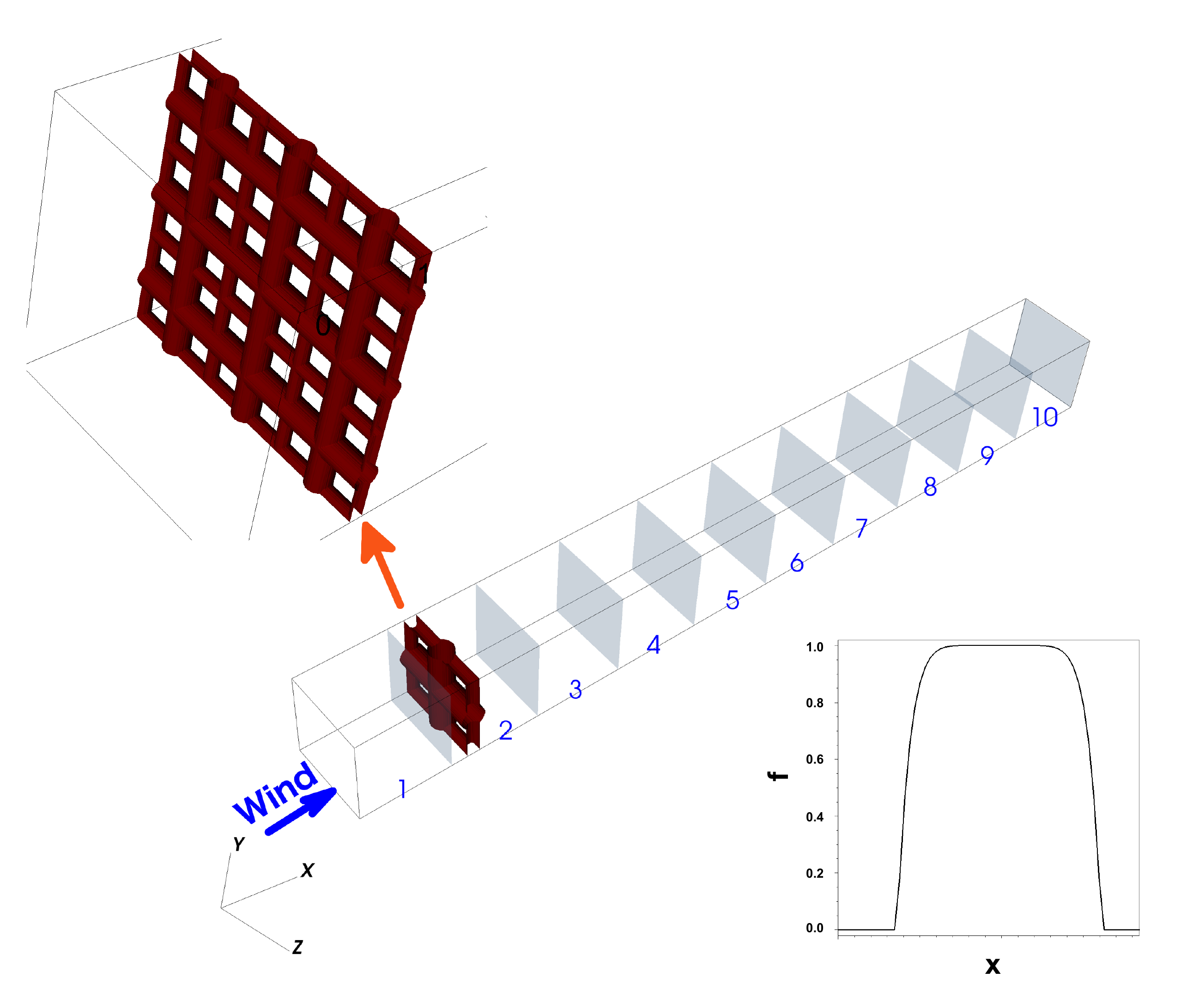}
%\caption{}
\caption{Simulation setup. A hypersonic wind was injected into the lower $x$ boundary of the computational domain. This wind collided with an embedded wire mesh at $x=1.25$. For clarity, an enlarged illustration of the mesh is shown in the upper left portion of the figure, which shows the relative spacing of the component wires and their diameters. Note, the physical domain actually encompasses only $1/9$ of the grating shown in this inset, however, given the periodic boundary conditions on the $y$ and $z$ faces of the box, a larger segment of the mesh is effectively modeled. The numbers on the shock tube ($\tilde{x}=1-10$), and corresponding gray planes, delineate the different sub-regions of the domain used in the analyses of Sections \ref{section: variances}-\ref{section:mach_estimate}. The lower right inset of the figure shows the smoothing function applied to the wires (see text for functional form). 
 }  
\label{fig:schematic}
\end{figure*}

The wires and ambient gas were initialized to be in pressure equilibrium at the start of the simulation in order to minimize expansion of the wires into the surrounding gas. Correspondingly, the wires were a factor of $10^4$ colder than the ambient gas. This setup allowed the gas to remain at rest until passage of the wind. The wire mesh was composed of a 2D lattice of cylindrical wires alternating in radius between $r=0.0625$ and $r=.03125$, located at $x=1.25$ (note, the entire computational volume was $10\times1\times1$ in $x$, $y$, and $z$, see Fig. \ref{fig:schematic}).  The different wire widths were used to impose multiple wavelength perturbations to the flow. The interwire spacing of the mesh was $\Delta=.40625$. A smoothing function of the form $f=\max[0,1-(x^2+y^2)^4]$ was applied to the wires to avoid sharp discontinuities in the fluid variables between the wires and ambient medium (c.f. lower-right inset of Fig. \ref{fig:schematic}). 

At $t=0$, a hypersonic post-shock wind (corresponding to a planar shock of sonic Mach number $M_s=20$, with respect to the pre-shock \textit{ambient gas}) was continuously injected into the lower $x$ boundary of the computational domain ($x=0$). Once the leading edge of this wind reached the upper $x$ boundary ($x=10$), the given simulation was terminated. Note, a magnetic field was injected into the domain along with the wind in the MHD case (i.e. only the wind contained a magnetic field, pre-shock gas did not). The injected field was oriented perpendicular to the direction of the wind ($\vec{B} = B\hat{y})$. The magnetic pressure in the wind ($P_{mag}\equiv \frac{B^2}{\sqrt{4\pi \rho_{wind}}}$) was ten times less than its thermal pressure (i.e. $\beta \equiv \frac{P_{gas}}{P_{mag}}=10$). Given these parameters, the wind was \textit{super-Alfv\'enic}, with an Alfv\'en Mach number $M_A\equiv(0.5\gamma\beta M_s^2)^{1/2}\approx 45$.

Both simulations were performed using upwinded Godunov-type integration methods that were second-order accurate in space and time. This included the piece-wise parabolic method (PPM) of \cite{colella1984} for the spatial reconstruction, % with a BLANK flux limiter, 
combined with the Harten-Lax-van Leer Contact (HLLC) Riemann Solver \citep{toro1994} for the hydro flux update and the HLLD Riemann solver for the MHD flux update \citep{miyoshi2005}. The magnetic field was evolved according to a corner transport upwind (CTU) scheme \citep{cunningham2009}. Time stepping was performed using a two-step Runge-Kutta method  \citep{shu1988}. 

The computational domain had a base grid composed of $1600x160^2$ computing zones. On top of this base grid, the simulations were initialized with $1$ level of refinement, which was centered on the wire mesh. Over the course of the simulation, gradients in the fluid variables triggered refinement in other regions of the computational domain. With the chosen refinement criteria, turbulent substructure was adequately captured at the highest resolution of the simulation, which was $\Delta x_{min}=3.13\times10^{-3}$.  Periodic boundary conditions were imposed on the $x-y$ and $x-z$ planes to enforce slab symmetry of the setup. Inflow/outflow boundaries were used on the remaining faces of the domain, corresponding to the lower/upper $y-z$ planes, respectively. 

\section{Results} \label{section: results}

In the following section, we present results of velocity and magnetic field dispersion along the length of the shock tube (Section \ref{section: variances}), power spectra of the velocity and magnetic fields (Section \ref{sec: spectra}), and probability density function fitting of the gas density (Section \ref{section:mach_estimate}). 
We begin by orienting the reader to the general morphological features of supersonic flows past a wire mesh revealed in our study.

\subsection{Flow Morphology}\label{section:morphology}

\begin{figure*}
\centering
\includegraphics[width=\textwidth]{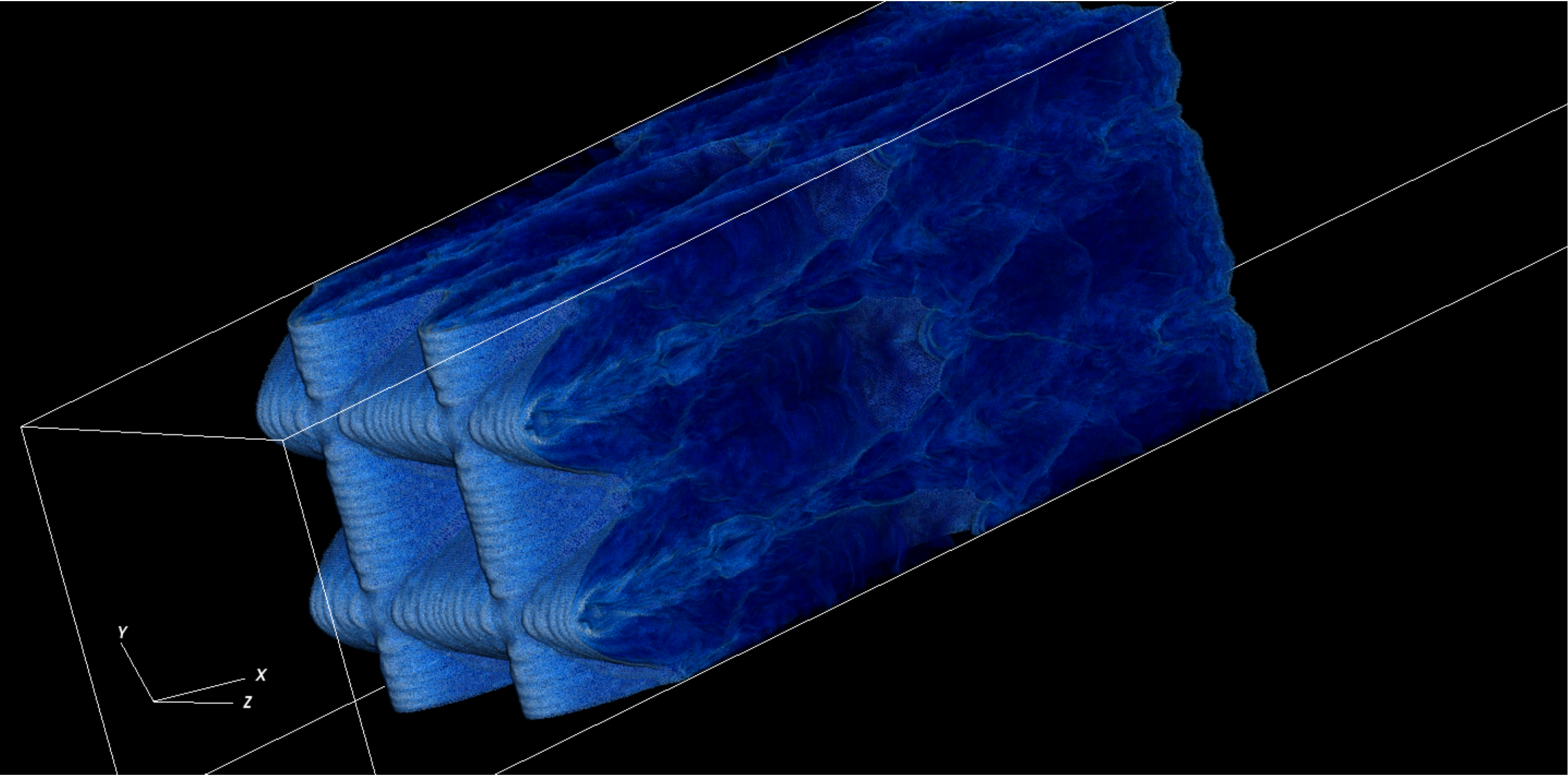}
\caption{Volume rendering for the hydrodynamic case. This plot illustrates the typical 3D structure generated by supersonic flow through a wire mesh. Intersecting bow shocks are visible outlining the constituent wires, which are being continuously ablated by the injected hypersonic flow (from the lower $x$ boundary). Kelvin-Helmholtz and Nonlinear Thin Shell instabilities can be seen in the bow shock layer and downstream gas arising from perturbations in the flow as it is shocked and deflected around the wires.} 
\label{fig:3d}
\end{figure*}

The volume rendering in Figure \ref{fig:3d} illustrates typical flow structures found in the simulations. First, intersecting bow shocks are formed along the face of the wire mesh as the injected hypersonic wind collides with wire material or passes through the mesh openings. The flow is strongly sheared as it passes through the bow shocks that wrap around the wires. The shocked flow shows excitation of Kelvin-Helmholtz (KH) instabilities along the shear layer. In addition to KH modes, the thin, isothermal shocks are unstable to Richtmeyer-Meshkov (RM) and Nonlinear Thin Shell (NTS; Vishniac 1994\nocite{vishniac1994}) instabilities. Together, these unstable modes seed perturbations that grow and propagate downstream. Thus, passage of the shock through the wire mesh transforms an initially laminar flow into one rich in multiscale, inhomogeneous structure.

\begin{figure*}
\centering
\includegraphics[width=\textwidth]{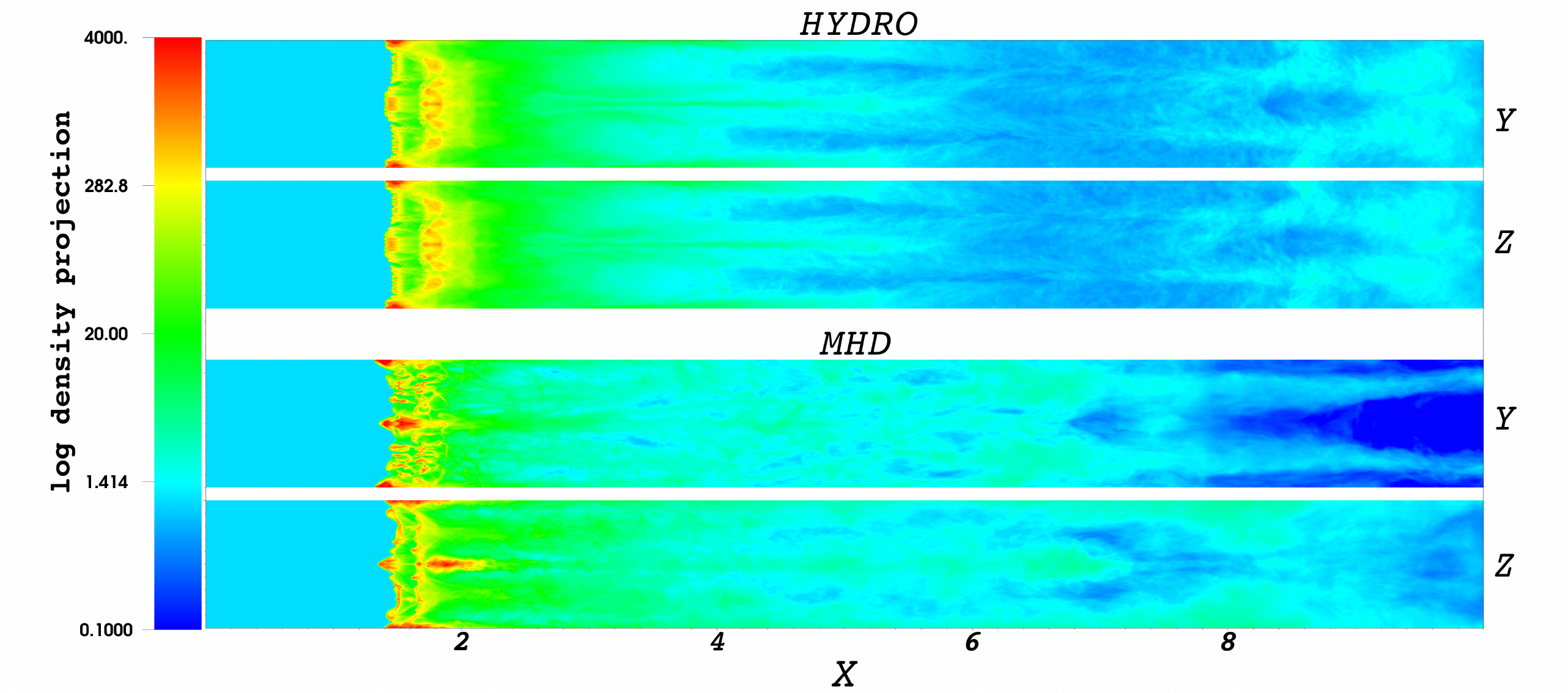}
\caption{Column density maps in the $x-z$ and $x-y$ planes for the hydro (top panel) and MHD (bottom panel) runs. Note, the color legend gives the density in $\log_{10}$ space. Dense knots and tendrils can be seen to form behind the wires ($x\approx 2$) as gas is shocked at the intersection points of multiple bow shocks. As described in the text, the dense features at the location of the wire mesh ($x=1.5$) are due to the integration of density through the wires and can be ignored. Turbulent sub-structure can be seen along the length of the shock tube as the flow passes through and ablates the wires. The initial orientation of the field before it passes through the wires ($\vec{B_0}=B_0\hat{y}$) breaks the symmetry between the different planes in the MHD case and results in overall higher densities.}
\label{fig:proj}
\end{figure*}

Figure \ref{fig:proj} shows density projections perpendicular to the bulk flow direction for the hydro and MHD runs at the final simulation time. Beginning with the top panel of the figure for the hydro case, the flow looks virtually the same irrespective of projection angle. Additionally, the dark `knots' occurring at $x\approx 1.5$ in both the $x-y$ and $x-z$ planes represent the wires (c.f. upper inset of Fig. \ref{fig:schematic}). Just behind the wires at $x \approx 2$, dense post-shock wind material is visible at the nodes of intersecting shocks (i.e. Mach stems), which occur as the gas passes through the `cells' of the wire mesh and collides downstream. In a sense, each cell creates an expanding jet-like flow that interacts with its neighbor. This produces interaction regions that are filamentary in structure and oriented mainly along $x$. Finally, further downstream, the projected density decreases as the flow rebounds from the intersecting shocks and expands. Eventually multiscale, turbulent substructure develops along the remaining length of the shock tube.

The MHD run (bottom panel Fig. \ref{fig:proj}) looks nearly identical to the hydro case, with two important differences. First, the MHD case exhibits some variation in the flow between the $x-y$ and $x-z$ projected planes. Second, the initial density inhomogeneities formed just beyond the wire mesh are denser in the MHD run compared to the hydro run. The break in symmetry between the $x-y$ and $x-z$ planes is due to the magnetic field being initially orientated along $y$ ($\vec{B_0}=B_0 \hat{y}$). As material streams through the wires, the magnetic field is dragged around the wires leading to extended/elongated loops of field oriented in the $x$ direction. Where the extended loops of field from neighboring mesh cells interact, the gas is compressed to higher densities. Thus, the increased density of the nonlinear structures formed just downstream of the wire mesh, as well as the differences between the projection maps along $y$ and $z$, are due to the distortion of the field as it is dragged around and through the wires.

\subsection{Time-averaged Velocity and Magnetic Field Dispersion}\label{section: variances}

\begin{figure*}
\centering
\includegraphics[width=\textwidth]{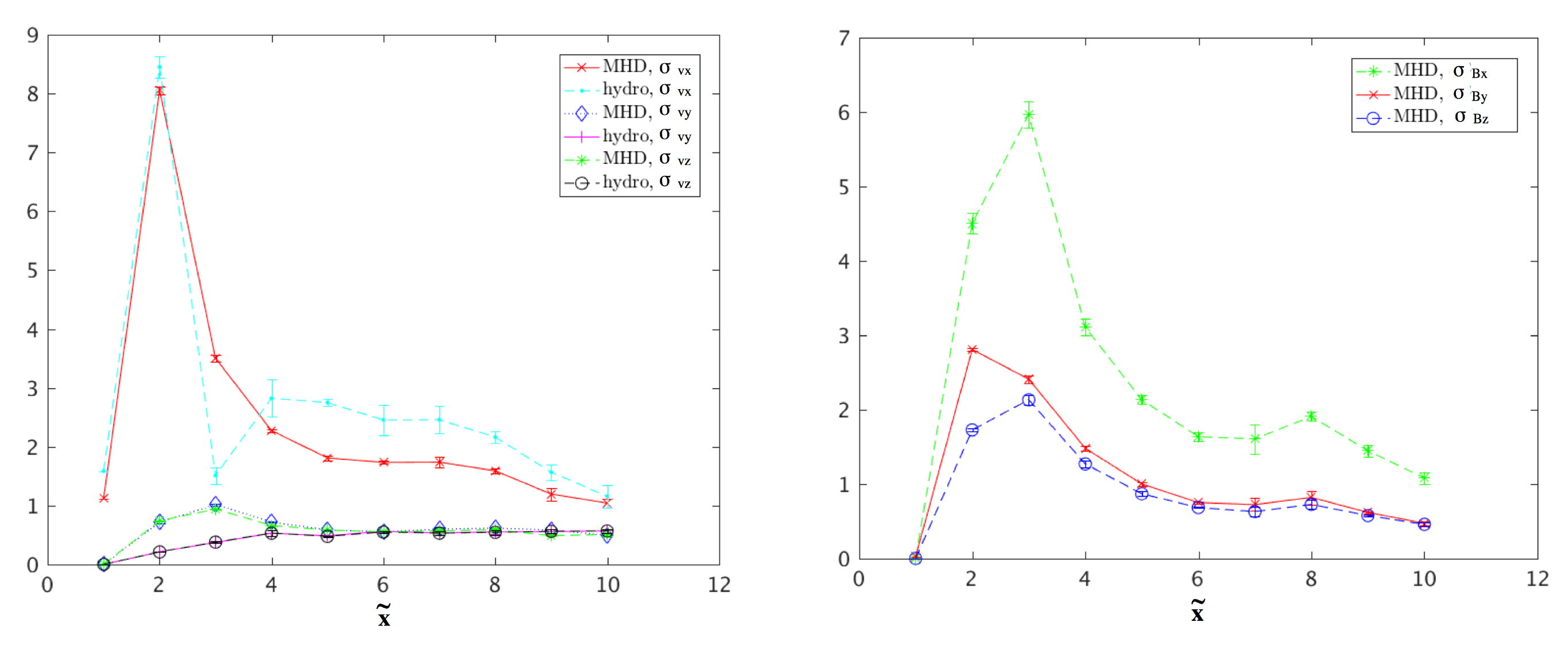}
\caption{Time-averaged standard deviation of velocity and magnetic field components along the shock tube. The $\tilde{x}$ axis gives the sub-region over which the given standard deviation was calculated and averaged (see text for details). \textit{Left panel} shows $\sigma_{v_x}$, $\sigma_{v_y}$, $\sigma_{v_z}$ for the hydro and MHD cases. The wires are contained within the analysis sub-region at $\tilde{x}=2$, hence the rise in $\sigma$ after this point for all components. However, given the flow is initially orientated along $x$ in both cases (perpendicular to the wire mesh), the largest increase in standard deviation occurs for $v_x$. In the MHD case, $\sigma_{v_x}$ is lower in the downstream gas ($\tilde{x}>4$) compared to the hydro simulation as the magnetic field resists gas motions in this direction. \textit{Right panel} shows the standard deviations of the magnetic field components for the MHD case. Given flux-freezing in ideal MHD, similar trends arise in the dispersion of the field components compared to velocity.} 
\label{fig:dev}
\end{figure*}

We now turn to more quantitative measures of the flow, beginning with velocity and magnetic field dispersion along the length of the shock tube. Standard deviations were calculated for each Cartesian component of the velocity and magnetic field in 10 sequential sub-regions of the domain. These were calculated over the last ten time states of the given simulation (where $\Delta t=0.005~t_{final}$), and averaged. The results are plotted in Figure \ref{fig:dev}.

Beginning with the velocity field (left hand panel, Fig. \ref{fig:dev}), the flow begins at $\tilde{x}=1$ oriented solely along $x$ ($\vec{v_0}=v_0 \hat{x}$), hence $\sigma_{v_y}=\sigma_{v_z}=0$ at $\tilde{x}=1$ (note, $v_x>>1$, so $\sigma_{v_x}\neq0$ at this position is due to numerical noise). At $\tilde{x}=2$, the wires are contained within the analysis sub-region. This accounts for the increase in the standard deviation of all velocity components as the flow travels through the wires and is deflected. However, the component perturbed \textit{ most strongly} upon passage through the wire mesh is $v_x$, evidenced by the sharp increase in $\sigma_{v_x}$ at this position. This result is consistent with the prominent elongated structures visible in the column density maps along $x$ (see Section \ref{section:morphology}). In the other coordinate directions, the velocity standard deviations are roughly isotropic.

Beyond the wires ($\tilde{x}>3$), the flow relaxes into more coherent velocity structures, visible by the steep decline in $\sigma_{v_x}$ and the flattening out of each of the curves. Yet, the velocity continues to be largely anisotropic, remaining preferentially perturbed along the bulk flow direction ($\sigma_{v_x} >\sigma_{v_y} \sim \sigma_{v_z} $). Further downstream still ($\tilde{x}>4$), $\sigma_{v_x}$  is \textit{lower} in the MHD case than in the hydro, which is consistent with the magnetic field resisting motions along this direction. Note, however, that only slight differences exist between the MHD and hydro runs for $\sigma_{v_y}$ and $\sigma_{v_z}$, which, once again, is consistent with the flow being only weakly perturbed in these directions as it passes through the wire mesh. Thus, the MHD velocity field is \textit{slightly} more isotropic downstream, given the restriction of fluid motions along $x$ by the magnetic field.

Turning now to the dispersion in the magnetic field (right hand panel, Fig. \ref{fig:dev}), note that the overall shape and trend seen in the curves are similar to those of the velocity standard deviations. This is not surprising given the field is perfectly coupled to the gas in ideal MHD (except on the grid scale, where numerical dissipation of the field can occur). Thus, magnetic field perturbations are tied to velocity motions and scatter is mirrored in each of the distributions. As described above for velocity, before the flow passes through the wire mesh at $\tilde{x}=1$, the magnetic field is in its initial, unperturbed state ($\vec{B_0}=B_0\hat{y}$). Once the flow drags the field through the wire mesh at $\tilde{x}=2$, a sharp increase is again visible in the standard deviation of the $x$ component of the field ($\sigma_{B_x} >> \sigma_{B_y} \sim \sigma_{B_z}$). The smaller, concomitant increases in $\sigma_{B_y}$ and $\sigma_{B_z}$ at this point are due to the flow being only marginally perturbed in $v_y$ and $v_z$, as described previously. Finally, far away from the wires ($\tilde{x}>2$), a decline in the dispersion of all components is again visible, with the standard deviation in $B_x$ remaining the largest downstream.

\subsection{Power Spectra}\label{sec: spectra}

\begin{figure}
\centering
\includegraphics[width=\columnwidth]{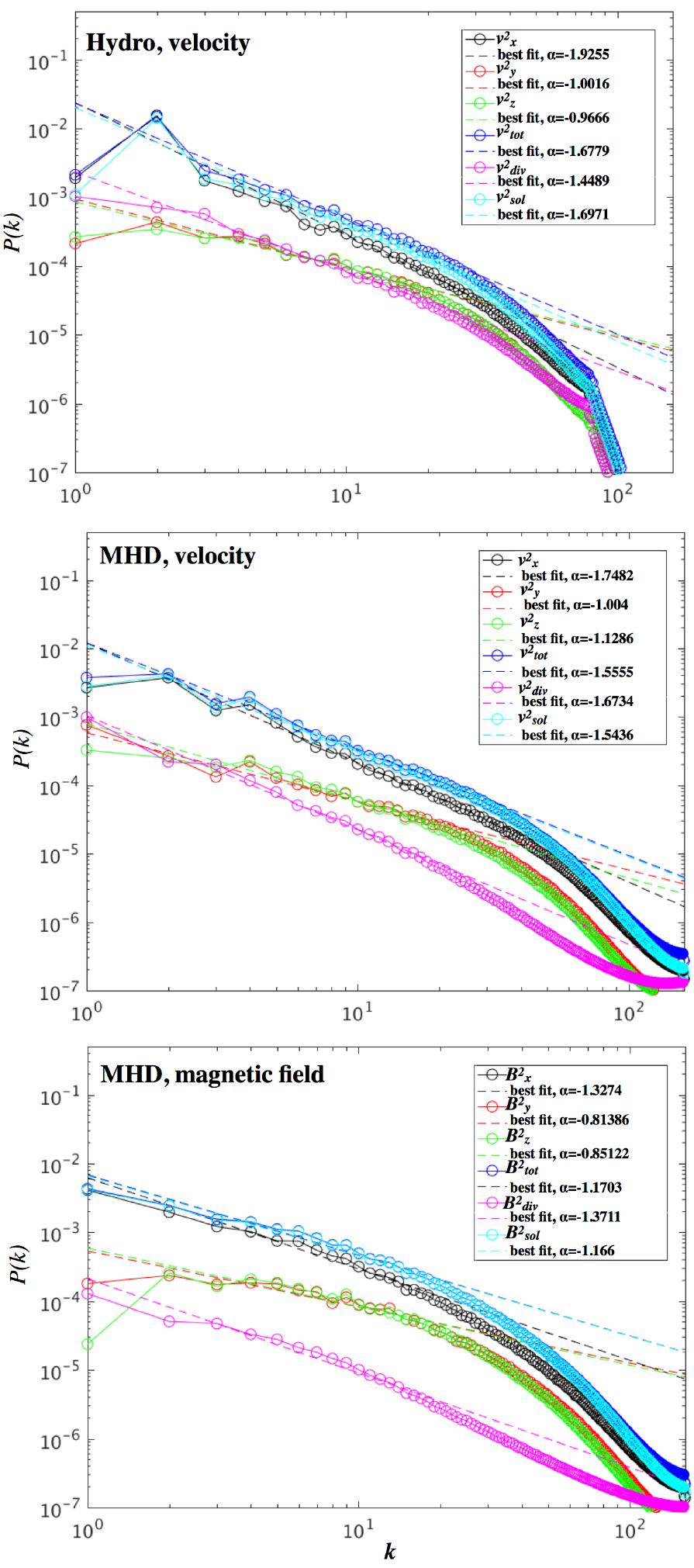}
\caption{Power spectra of the velocity and magnetic fields for the different runs. Power spectra were computed for the Cartesian and Helmholtz decomposed components of the fields, as described in the text. Best fit lines were calculated for each spectrum over the inertial range captured by the simulation ($2<k<20$), and are shown as dashed lines in the figure. The slopes of these best fit lines ($\alpha$) are given in the legend.} 
\label{fig:spectra}
\end{figure}

Power spectra were constructed for the Cartesian ($\hat{x}$, $\hat{y}$, $\hat{z}$) and Helmholtz-decomposed solenoidal and compressive components of the velocity and the magnetic fields (satisfying $\nabla \cdot \vec{a}=0$ and $\nabla \times \vec{a}=0$, respectively) %. Spectra were computed 
using AstroBEAR's discrete Fourier transform module, which performs fast Fourier transforms (FFT) of the data on an AMR mesh (see Carroll-Nellenback et al. 2014\nocite{carroll2014} for details). %In 1D, the FFT of a function $f(x)$ is given by $F(k)=\sum_{x=1}^{N}\mathrm{e}^{\frac{2 \mathrm{\pi i}}{N}k x}f(x)$, where N is the number of discrete grid points to sample the function and k is the wave number normalized to the length of the region over which the FFT is being computed ($k=2\mathrm{\pi}L^{-1}$). For a 3D function, $f(x,y,z)$, the FFT is computed over each dimension sequentially. By Parseval's theorem, the spectral power distribution $P(k)$ of $f$ is then given by multiplying each component of the FFT by its complex conjugate, i.e. $P(k)=F(k) \bar{F}(k)$. Thus, power spectra of a given quantity represent the square of that quantity in $k$-space. 
Spectra were calculated for the last time state in each of the runs, in the last region of the computational domain ($\tilde{x}=10$). % the last sub-domain of the computational box, $\tilde{x}=10$, 
Given the location of this region furthest away from the wire mesh, nonlinear perturbations seeded in the flow by the KH, RM, and NTS instabilities have had the most time to homogenize and decay into a turbulent cascade. Before discussing specific details about the power spectra for the different runs, we would like to make a few broad remarks about the trends seen in each of the panels of Figure \ref{fig:spectra}. First, note that the driving scale for energy injection is visible at $k \approx 2$, corresponding to a physical wavelength of $\lambda \approx .5$. % Note, I took out the over kmin factor given L = 1.
This length scale is consistent with the inter-wire spacing of the mesh, $\Delta \approx 0.4$. Second, at the high $k$ end of the spectrum, numerical dissipation on the grid scale limits the inertial range of any turbulent energy cascade that could be captured by the simulation. This is evidenced by the decline in power for each of the curves beyond $k>20$. Thus, we take the inertial range captured by the simulations to lie between $2<k <20$. Lastly, each spectrum exhibits a scaling law over this limited range in $k$ ($P\propto k^{\alpha}$). To quantify the slope of the power-law scaling in each of the cases, best-fit lines are overlaid in the figure. 

Turning now to the power spectra of velocity for the hydro case (leftmost panel of Figure \ref{fig:spectra}), the slope of the best-fit line to the $v_x^2$ spectrum is $\alpha = -1.9$, which is very near the value expected for a turbulent cascade under supersonic, compressible, isothermal conditions (i.e. $\alpha=-2$, see for example, Kritsuk et al. 2007; Schmidt et al. 2009\nocite{schmidt2009}\nocite{kritsuk2007}). By comparison, the slopes of the best fit lines in the other dimensions are $\alpha =-1.0$ and $\alpha=-.97$ for $v_y^2$ and $v_z^2$, respectively. That the slopes vary for the different components of $v$ indicate that specific kinetic energy was deposited to the flow \textit{anisotropically}. Moreover, the results support that the flow was most strongly perturbed along the direction of the wind propagation, since the overall power of the $v_x$ mode was greater than $v_y$ and $v_z$ on all scales.

Considering next the \textit{total} hydro velocity power spectrum, it may be unsurprising that the derived slope lies between $1<\alpha<2$, given the slopes of the individual components. However, it is interesting how close this value ($\alpha=-1.6779$) lies to the classical \cite{kolmogorov1941} spectrum for incompressible turbulence ($\alpha = -5/3$). Power spectrum of the Helmholtz-decomposed velocity field for the hydro case indicates that more energy was injected into solenoidal modes than compressive (i.e. $v^2_{sol}>v^2_{div} ~\forall~ k$). The ratio of solenoidal to compressive energy can be used to determine the effective sonic Mach number of a turbulent flow \citep{molina2012}, and is investigated below in Section \ref{section:mach_estimate}. 

Turning now to the MHD case, power spectra of the velocity (middle panel, Fig. \ref{fig:spectra}) show similar trends and scaling to those in the hydro simulation, with the largest difference being the slope of the compressive velocity mode, $v^2_{div}$, which has decreased to $\alpha=-1.67$. Consequently, \textit{even more energy} ($E_k=\int P_k dk$) has been deposited into solenoidal modes than compressive in the MHD case. Power spectra of the magnetic field components (right panel, Fig. \ref{fig:spectra}) mirror the MHD velocity spectra as expected, due to the ideal MHD nature of the simulations.

\subsection {Mach Number Estimation}\label{section:mach_estimate}

The distribution of gas density in simulations of isothermal, supersonic turbulence is well known to follow a log-normal probability density function (PDF) \citep{blaisdell1993, Vazquez-Semadeni1994, padoan1997, passot1998, scalo1998, kritsuk2007, federrath2012}. In the isothermal limit, it is often convenient to define the \textit{logarithmic} density, $s=\ln{(\rho/\rho_0)}$ (where $\rho_0$ is the mean gas density). This change of variables allows the log-normal PDF to be expressed as a simple Gaussian distribution in $s$:

\begin{equation}\label{eqn:pdf}
p(s)=\frac{1}{\sqrt{2\pi\sigma^{2}_{s}}}\exp\left(1-\frac{(s-s_{0})^2}{2\sigma^{2}_{s}}\right).
\end{equation}

\noindent Here, $\sigma_s$ denotes the standard deviation of $s$, and $s_0$ is the mean logarithmic density in the flow, as usual for the normal distribution. 

Early studies of hydrodynamic turbulence showed that increasing the RMS sonic Mach number ($M$) of the flow led to a proportional increase in the standard deviation of the gas density, i.e. $\sigma_\rho=b M$ (e.g. Padoan et al. 1997; Passot and V\'{a}zquez-Semadeni 1998\nocite{padoan1997, passot1998}). The proportionality constant $b$ in this relation was later shown to depend on the \textit{type} of turbulent modes present in the flow, ranging from $b=1/3$ for purely solenoidal to $b=1$ for compressive \citep{Federrath08, Federrath10}.  %seifried2011}. 
For this reason, $b$ is today commonly referred to as the `turbulent forcing parameter'. 

Assuming the logarithmic gas density follows a Gaussian PDF (Eqn. \ref{eqn:pdf}), it can be shown for \textit{hydrodynamic} turbulence that $\sigma_s$, $b$, and $M$ are related through $\sigma_s^2=\ln({1+b^2M^2})$ \citep{padoan1997}. For MHD turbulence, this formula must be corrected to account for the magnetic field. \cite{molina2012} derived the following semi-analytical formula for $\sigma_s$ (again assuming a Gaussian PDF in $s$), as a function of $b$, RMS sonic Mach number $M$, and $\beta$ 

\begin{equation}
\sigma_{s}^{2}=\ln\left(1+b^{2}M^{2}\frac{\beta}{\beta+1} \right).
\label{eq:machb}
\end{equation}

\noindent Equation \ref{eq:machb} has the necessary property that it reduces to the hydrodynamic formula in the limit ($\beta\rightarrow \infty$), and agrees well with 3D simulations of super-Alfv\'{e}nic MHD turbulence \citep{molina2012}. 

\begin{figure*}
\centering
\includegraphics[width=\textwidth]{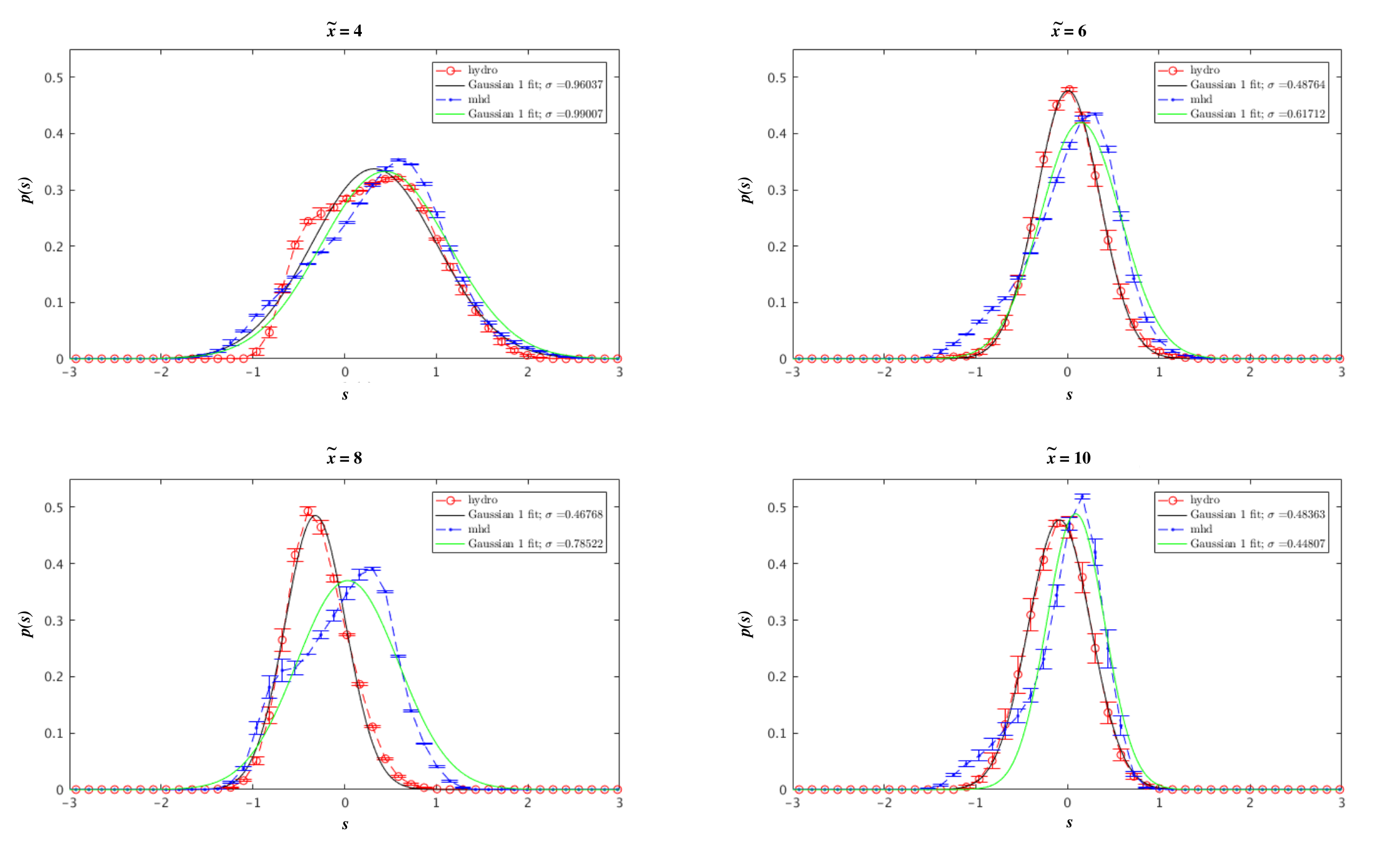}
\caption{Volume-weighted density PDFs and corresponding best fit Gaussian functions. \textit{Counter clockwise from left}, PDFs are computed within the sub-regions $\tilde{x}=4$, $6$, $8$, $10$, and averaged over the last ten time states of the simulation (where $\Delta t=0.005~t_{final}$). Red and blue curves correspond to hydro and MHD data, respectively, and error bars are overlaid on the curves representing $\pm$ one standard deviation over the average. Black and green curves correspond to the respective best fit Gaussian functions. The standard deviations of these Gaussians ($\sigma_s$) are listed in the legend, and used in Equation \ref{eq:machb} to calculate the estimated RMS sonic Mach number for the runs along the length of the shock tube (c.f. Fig. \ref{fig:mach}).     
 }
\label{fig:rhoPDF}
\end{figure*}

Using Equation \ref{eq:machb}, one can estimate $M$ of a turbulent flow, given suitable approximations of $\sigma_s$, $b$, and $\beta$. To approximate $\sigma_s$ for the different runs, we fit Gaussian distributions to volume-weighted density PDFs of the data (Fig. \ref{fig:rhoPDF}). The PDFs were constructed for each of the sub-regions along the shock tube, and averaged over the last ten time states of the simulation. These are shown clockwise in the figure for $\tilde{x}=4$, $6$, $8$, and $10$ (blue points corresponding to hydro data and red to MHD). Best fit Gaussians are overlaid in each of the panels (solid black and green curves for the hydro and MHD fits, respectively). %The standard deviations of these best fit Gaussians are given in the legend. 
As can be seen in the figure, Gaussian distributions closely approximate the PDFs, and the dispersion of the data is minimal over the last ten time states. Thus, we use the standard deviations of these best fit Gaussians (given in the legend) in Equation \ref{eq:machb} to compute the estimated $M$ as described below.

Before discussing our best estimate of $M$ in each of the cases, we draw the readers attention to the evolution of the PDFs shown in Figure \ref{fig:rhoPDF}. Note that the shapes of the PDFs change in traversing the computational domain from $\tilde{x}=4$ to $\tilde{x}=10$; both the hydro and MHD PDFs become more tightly peaked for larger $\tilde{x}$. \cite{Federrath08, Federrath10} showed that the width of density PDFs in simulations of fully developed turbulence correlates with the relative strength of compressive vs. solenoidal modes in the flow, with solenoidal flows producing narrower distributions than compressive (see for example fig. 2 and 6 in those papers). Thus, these results might suggest that the flow switches from being more strongly compressive in regions closer to the wires, to more solenoidal in regions further downstream. This behavior would be consistent with \cite{Federrath10}, which shows that even turbulence driven by purely compressive forcing can decay into some fraction of solenoidal modes (about $1/2$ for 3D flows). Power spectra of the flows also supported that the velocity field was predominantly solenoidal at large $\tilde{x}$ (Section \ref{fig:spectra}).

\begin{figure*}
\centering
\includegraphics[width=0.7\textwidth]{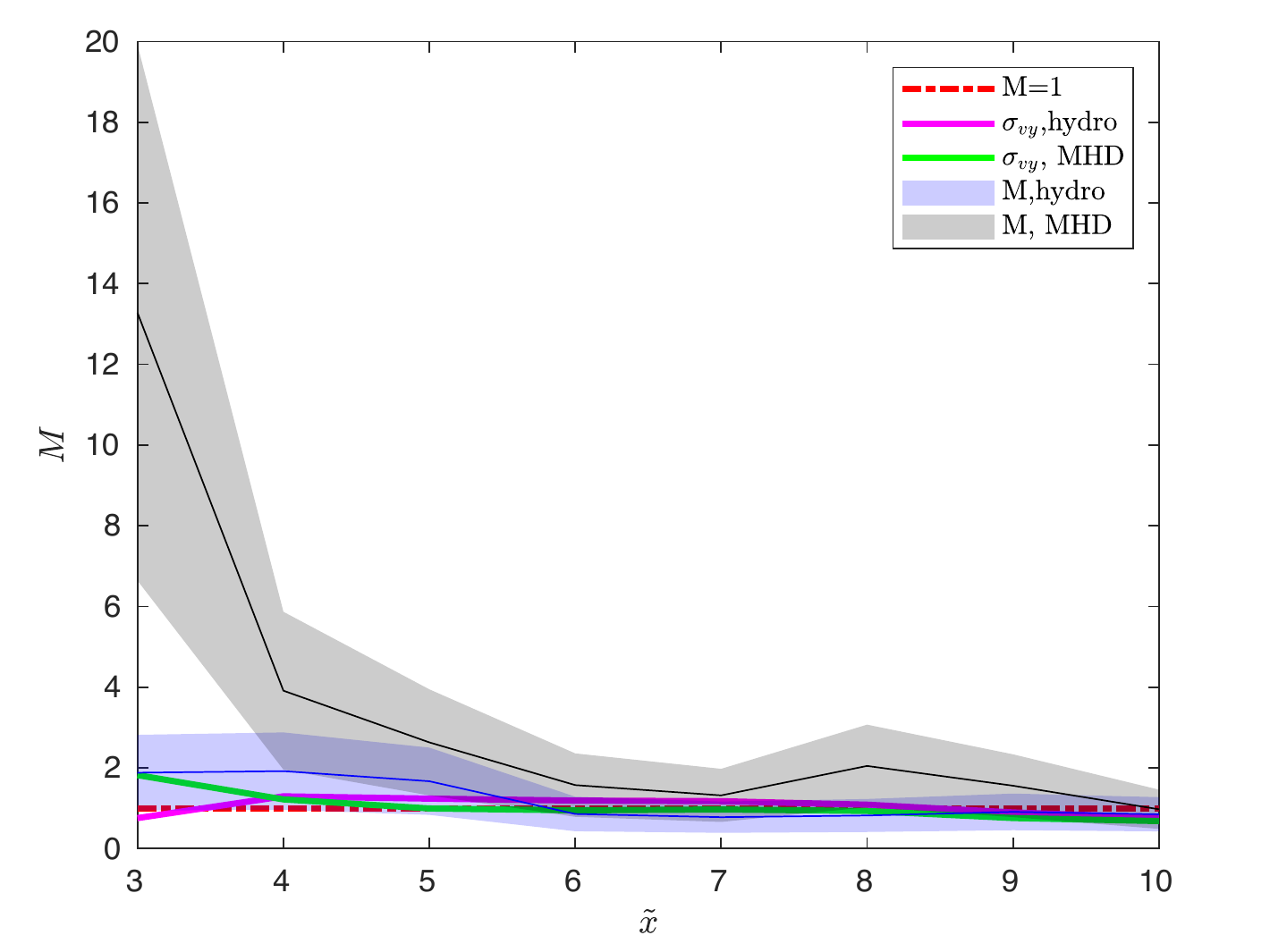}
\caption{Estimated RMS sonic Mach number ($M$) along the length of the shock tube for the different runs. Estimates of $M$ were computed along $\tilde{x}$ using Equation \ref{eq:machb} with the respective best fit of $\sigma_s$ for each of the density distributions (see Fig. \ref{fig:rhoPDF}), $\beta=\infty$ for the hydro simulation, and a spatially and temporally averaged value of $\beta$ in the MHD simulation. The estimated $M$ for the hydro run lies along the blue curve, and along the black for the MHD run. A range of $M$ is plotted (given by the shaded region of corresponding color in the figure), accounting for the uncertainty in the turbulent forcing parameter $b$ (which has previously been shown to vary between $1/3 \leq b \leq 1$, see text for details). The standard deviation of a perpendicular velocity component with respect to the bulk flow ($\sigma_{v_y}$) is overlaid on the plot for comparison (recall $M\approx v_{rms}$ for isothermal flow). As can be seen in the figure, the flow is largely transonic downstream of the wires in both cases, with $M\approx1$ at $\tilde{x}=10$.}
\label{fig:mach}
\end{figure*}

We now turn to Figure \ref{fig:mach}, which shows the estimated RMS sonic Mach number $M$ along the shock tube in both the hydro and MHD cases. For each case, we used a spatially averaged value for  $\beta$ in each of the sub-domains of the shock tube ($\tilde{x}= 1-10$), averaged over the last ten time states. For both the hydro (blue regions of the plot) and MHD data (black regions of the plot), a range in potential turbulent forcing parameter $b$ is considered ($1/3\leq b\leq1$) in calculating $M$ for a given $\tilde{x}$. This is represented in the figure as the shadowed area above and below the solid curves of the same color (the intermediate value of $b$ lies along the solid curve). Thus, for a given $\tilde{x}$, the highest possible $M$ corresponds to $b=1/3$, and the lowest $b=1$. The range of possible $M$ also depends on the best fit $\sigma_s$, which varies along $\tilde{x}$ as described previously. As can be seen in the figure, \textit{at best} ($b=1/3$), $M$ is \textit{marginally} supersonic for most of the domain beyond the wires ($\tilde{x}>5$) in both the hydro and MHD cases, and transonic in the longest evolving regions of the flow, i.e. $M\approx1$ at $\tilde{x}=10$ (note that shock compression of the field immediately beyond the wires led to a steep decrease in $\beta$ in that region, thus explaining the high mach numbers immediately adjacent to the wires, i.e. $\tilde{x}= 3$).

\section{Discussion}\label{section: discussion}

We have presented a set of high-resolution, 3D AMR simulations to test whether supersonic MHD turbulence could be generated by passing a hypersonic ($M_s=20$), super-Alfv\'{e}nic ($M_A=45$) flow through a wire mesh. Our findings support that \textit{marginally} supersonic turbulence can be achieved by this experimental setup, irrespective of upstream hydrodynamic or MHD conditions. Further, our results are strongly consistent with the generation of \textit{anisotropic} turbulence in the downstream (post-shock) flow. We presented a number of analyses in support of these conclusions, including a study of the morphological features of the flows (Section \ref{section:morphology}), time-averaged dispersion of the velocity and magnetic field along the shock tube (Section \ref{section: variances}), power spectra of the velocity and magnetic fields (Section \ref{sec: spectra}), and a sonic Mach number estimate based on the post-shock gas density PDF for each of the runs (Section \ref{section:mach_estimate}). 

The prominent morphological features of hypersonic flow past a wire mesh include unstable bow shocks surrounding the wires and multiscale post-shock turbulent substructure oriented mainly along the bulk flow direction ($\hat{x}$). These structures arise as follows. In both the hydro and MHD cases, the incident hypersonic wind is shocked and deflected around the wires. This leads to the excitation of various fluid instabilities in the bow shock layer (such as the KH, RM, and NTS instabilities), which causes ripples and distortions along the bow shock front. This effect is clearly visible in the volume rendering of Figure \ref{fig:3d}, but is also apparent when viewed in projection (Fig. \ref{fig:proj}). As each bow shock wraps around its associated wire, the interaction of bow shocks from neighboring wires then leads to the formation of mach stems in the region immediately downstream of the wire mesh. Together, these shock intersections and their associated instabilities seed density perturbations which grow into inhomogeneous clumpy and filamentary substructure further downstream. Throughout the the simulation, the density inhomogeneities remain oriented predominantly along the bulk flow direction in both the hydro and MHD cases.

The standard deviations of the velocity field ($\sigma_{vx}$, $\sigma_{vy}$, $\sigma_{vz}$) reflect the development of the morphological structures seen in the simulation. As shown in Figure \ref{fig:dev}, the component of the flow perturbed the most upon passage through the wire mesh was $v_x$, evidenced by a sharp increase in $\sigma_{v_x}$ in the region of the shock tube containing the wires. Further, this component remained the most variable along the remainder of the shock tube, indicative of perturbations being predominantly amplified in the $x$ direction. These trends were largely the same in both the hydro and MHD cases, with only minor differences arising between the runs. The dispersion of the magnetic field for the MHD case was also presented in Figure \ref{fig:dev} (right hand panel), with the trends mirroring those found in velocity, as to be expected due to flux-freezing in ideal MHD.  

Power spectra were presented of the velocity and magnetic field at the end of the simulation showing a power-law scaling relation in wave number (i.e. $P\propto k^{-\alpha}$), characteristic of turbulent flows. While the scaling power $\alpha$ varied between $1\lesssim \alpha\lesssim 2$ for the different components, they remained comparable between the hydro and MHD cases. In both runs, however, the component closest to a supersonic Burgers-type cascade ($\alpha=2$) was $v^2_x$. This supports that turbulence was most strongly seeded along the initial flow direction.

PDFs of the gas density indicated a relatively stronger solenoidal component of the flow compared to compressive in the regions furthest away from the wire mesh (Fig. \ref{fig:rhoPDF}). This finding was somewhat qualitative; \cite{Federrath08, Federrath10} showed that the width of the gas density PDF for purely solenoidally driven turbulence is significantly narrower than purely compressive turbulence of the same $v_{rms}$ and $M$. In traversing the shock tube, we found the widths (at half maximum) %the maximum of the respective distribution) 
for both the hydro and MHD gas density PDFs to decrease by roughly $50\%$, suggesting that the flow started out more compressive in regions nearest the wire grate, but became more solenoidally-dominated downstream. This could be the result of energy transfer between the modes by nonlinear interactions in the flow. For instance, \cite{Federrath10} showed that energy transfer between compressive and solenoidal modes readily occurs in high resolution 3D turbulence simulations. As shown in that paper, even turbulence driven by fully compressive forcing decays into a mixture of compressive and solenoidal velocity modes, with the ratio being $\approx 1:1$. 

Given that our gas density followed a nearly log-normal distribution, in agreement with many previous studies of supersonic, isothermal turbulence, e.g. \cite{blaisdell1993, passot1998, kritsuk2007}, we were able to estimate the RMS sonic Mach number of the flows according to Equation \ref{eq:machb}. We found $M$ to lie between $1\lesssim M\lesssim3$ for \textit{most} of the downstream shock tube (in both the hydro and MHD cases). In the region furthest away from the wire mesh, the range in probable $M$ was found to relax to transonic values ($M\approx 1$ for $\tilde{x}=10$).

We would like to finish this discussion with a few remarks on our choice of wire initialization. By treating the wires as an overdense fluid rather than a fixed, solid boundary, some ablation of wire material naturally resulted as the shock interacted with the mesh. In the present simulations, roughly $30 \%$ of the downstream gas (located at $\tilde{x}=10$) was ablated wire material. In the current setup, we expect the dominant effect of this wire ablation and entrainment would have been the seeding of turbulence, given the relative temperatures of the two phases of the gas (recall the wire gas was $10^4$ times colder than the surrounding post-shock gas which constituted the majority of the shock tube). The difference in sound speeds of these two materials alone would have translated into the wire gas being insufficient in driving significant thermal dynamics in the ambient medium. 

While we did not investigate the effect of wire ablation and entrainment on turbulence generation in the present set of simulations, it would be very interesting to explore this in future work. This is especially true since at least some degree of ablation could occur experimentally. While ablated mass in experiments such as \cite{Burdiak2017} would probably have been small on the time-scale of the experiment, the situation could be quite different in the case where the wires are ``preconditioned," e.g. by a separate current pulse, and converted into a lower density vapor state. To our knowledge, no physical experiments have been done on this as yet, but we would expect considerable mass entrainment in those cases. One way of varying the degree of wire ablation numerically would be to vary the density contrast between the wires and the ambient medium. Experimentally, this could be done using different materials for the plasma flow and the wire obstacle (solid or with reduced density), and then use spectroscopic diagnostics to observe the presence of the obstacle material in the post-wire plasma flow.

\section{Conclusions}

Our results suggest that experiments using a shock-tube like configuration with a wire mesh can, in fact, produce turbulent magnetized flows. However, because the turbulence requires passage through an obstacle (the wire mesh), the downstream flows remain close to sonic. Thus, any experiments using this configuration must be tuned to astrophysical environments dominated by such transonic flows, such as those in galaxy clusters.  The strongly supersonic turbulence found in star formation regions likely can not be explored using the configuration presented in this paper.

\section*{Acknowledgments}

This work used the computational and visualization resources in the Center for Integrated Research Computing (CIRC) at the University of Rochester and the computational resources of the Texas Advanced Computing Center (TACC) at The University of Texas at Austin, provided through allocation TG-AST120060 from the Extreme Science and Engineering Discovery Environment (XSEDE), which is supported by National Science Foundation grant number ACI-1548562. Financial support for this project was provided by the Department of Energy grant DE-SC0001063, the National Science Foundation grants AST-1515648 and AST-1813298, and the Space Telescope Science Institute grant HST-AR-12832.01-A. This work was also supported by the US Department of Energy through the Los Alamos National Laboratory. Los Alamos National Laboratory is operated by Triad National Security, LLC, for the National Nuclear Security Administration (Contract No. 89233218CNA000001). 

%\section{Conclusions}
%The last numbered section should briefly summarise what has been done, and describe the final conclusions which the authors draw from their work.

\bibliographystyle{mnras}
\bibliography{WT} 

\newcommand{\noop}[1]{}
\begin{thebibliography}{}
\makeatletter
\relax
\def\mn@urlcharsother{\let\do\@makeother \do\$\do\&\do\#\do\^\do\_\do\%\do\~}
\def\mn@doi{\begingroup\mn@urlcharsother \@ifnextchar [ {\mn@doi@}
  {\mn@doi@[]}}
\def\mn@doi@[#1]#2{\def\@tempa{#1}\ifx\@tempa\@empty \href
  {http://dx.doi.org/#2} {doi:#2}\else \href {http://dx.doi.org/#2} {#1}\fi
  \endgroup}
\def\mn@eprint#1#2{\mn@eprint@#1:#2::\@nil}
\def\mn@eprint@arXiv#1{\href {http://arxiv.org/abs/#1} {{\tt arXiv:#1}}}
\def\mn@eprint@dblp#1{\href {http://dblp.uni-trier.de/rec/bibtex/#1.xml}
  {dblp:#1}}
\def\mn@eprint@#1:#2:#3:#4\@nil{\def\@tempa {#1}\def\@tempb {#2}\def\@tempc
  {#3}\ifx \@tempc \@empty \let \@tempc \@tempb \let \@tempb \@tempa \fi \ifx
  \@tempb \@empty \def\@tempb {arXiv}\fi \@ifundefined
  {mn@eprint@\@tempb}{\@tempb:\@tempc}{\expandafter \expandafter \csname
  mn@eprint@\@tempb\endcsname \expandafter{\@tempc}}}

\bibitem[\protect\citeauthoryear{{Alexandrova}, {Carbone}, {Veltri}  \&
  {Sorriso- Valvo}}{{Alexandrova} et~al.}{2008}]{Alexandrova2008}
{Alexandrova} O.,  {Carbone} V.,  {Veltri} P.,   {Sorriso- Valvo} L.,  2008,
  \mn@doi [\apj] {10.1086/524056}, \href
  {https://ui.adsabs.harvard.edu/\#abs/2008ApJ...674.1153A} {674, 1153}

\bibitem[\protect\citeauthoryear{{Balsara}, {Benjamin}  \& {Cox}}{{Balsara}
  et~al.}{2001}]{balsara2001}
{Balsara} D.,  {Benjamin} R.~A.,   {Cox} D.~P.,  2001, \mn@doi [\apj]
  {10.1086/323967}, \href
  {https://ui.adsabs.harvard.edu/\#abs/2001ApJ...563..800B} {563, 800}

\bibitem[\protect\citeauthoryear{{Bellan} et~al.,}{{Bellan}
  et~al.}{2009}]{2009PhPl...16d1005B}
{Bellan} P.~M.,  et~al., 2009, \mn@doi [Physics of Plasmas]
  {10.1063/1.3101812}, \href
  {http://adsabs.harvard.edu/abs/2009PhPl...16d1005B} {16, 041005}

\bibitem[\protect\citeauthoryear{{Blaisdell}, {Mansour}  \&
  {Reynolds}}{{Blaisdell} et~al.}{1993}]{blaisdell1993}
{Blaisdell} G.~A.,  {Mansour} N.~N.,   {Reynolds} W.~C.,  1993, \mn@doi
  [Journal of Fluid Mechanics] {10.1017/S0022112093002848}, \href
  {http://adsabs.harvard.edu/abs/1993JFM...256..443B} {256, 443}

\bibitem[\protect\citeauthoryear{{Bott-Suzuki} et~al.,}{{Bott-Suzuki}
  et~al.}{2015}]{Bott-Suzuki2015}
{Bott-Suzuki} S.,  et~al., 2015, \mn@doi [Physics of Plasmas]
  {https://doi.org/10.1063/1.4921735}, 22, 052710

\bibitem[\protect\citeauthoryear{{Bruno} \& {Carbone}}{{Bruno} \&
  {Carbone}}{2013}]{bruno2013}
{Bruno} R.,  {Carbone} V.,  2013, \mn@doi [Living Reviews in Solar Physics]
  {10.12942/lrsp-2013-2}, \href
  {http://adsabs.harvard.edu/abs/2013LRSP...10....2B} {10, 2}

\bibitem[\protect\citeauthoryear{{Burdiak} et~al.,}{{Burdiak}
  et~al.}{2017}]{Burdiak2017}
{Burdiak} G.~C.,  et~al., 2017, \mn@doi [Physics of Plasmas]
  {10.1063/1.4993187}, \href
  {http://adsabs.harvard.edu/abs/2017PhPl...24g2713B} {24, 072713}

\bibitem[\protect\citeauthoryear{Carroll-Nellenback, Shroyer, Frank  \&
  Ding}{Carroll-Nellenback et~al.}{2013}]{carroll13}
Carroll-Nellenback J.~J.,  Shroyer B.,  Frank A.,   Ding C.,  2013, Journal of
  Computational Physics, 236, 461

\bibitem[\protect\citeauthoryear{{Carroll-Nellenback}, {Frank}  \&
  {Heitsch}}{{Carroll-Nellenback} et~al.}{2014}]{carroll2014}
{Carroll-Nellenback} J.~J.,  {Frank} A.,   {Heitsch} F.,  2014, \mn@doi [\apj]
  {10.1088/0004-637X/790/1/37}, \href
  {http://adsabs.harvard.edu/abs/2014ApJ...790...37C} {790, 37}

\bibitem[\protect\citeauthoryear{{Ciardi} et~al.,}{{Ciardi}
  et~al.}{2007}]{2007PhPl...14e6501C}
{Ciardi} A.,  et~al., 2007, \mn@doi [Physics of Plasmas] {10.1063/1.2436479},
  \href {http://adsabs.harvard.edu/abs/2007PhPl...14e6501C} {14, 056501}

\bibitem[\protect\citeauthoryear{{Colella} \& {Woodward}}{{Colella} \&
  {Woodward}}{1984}]{colella1984}
{Colella} P.,  {Woodward} P.~R.,  1984, \mn@doi [Journal of Computational
  Physics] {10.1016/0021-9991(84)90143-8}, \href
  {http://adsabs.harvard.edu/abs/1984JCoPh..54..174C} {54, 174}

\bibitem[\protect\citeauthoryear{{Couch} \& {Ott}}{{Couch} \&
  {Ott}}{2015}]{couch2015}
{Couch} S.~M.,  {Ott} C.~D.,  2015, \mn@doi [\apj] {10.1088/0004-637X/799/1/5},
  \href {https://ui.adsabs.harvard.edu/\#abs/2015ApJ...799....5C} {799, 5}

\bibitem[\protect\citeauthoryear{{Cunningham}, {Frank}, {Varni{\`e}re},
  {Mitran}  \& {Jones}}{{Cunningham} et~al.}{2009}]{cunningham2009}
{Cunningham} A.~J.,  {Frank} A.,  {Varni{\`e}re} P.,  {Mitran} S.,   {Jones}
  T.~W.,  2009, \mn@doi [\apjs] {10.1088/0067-0049/182/2/519}, \href
  {http://adsabs.harvard.edu/abs/2009ApJS..182..519C} {182, 519}

\bibitem[\protect\citeauthoryear{{Drake} et~al.,}{{Drake}
  et~al.}{2002}]{2002ApJ...564..896D}
{Drake} R.~P.,  et~al., 2002, \mn@doi [\apj] {10.1086/324194}, \href
  {http://adsabs.harvard.edu/abs/2002ApJ...564..896D} {564, 896}

\bibitem[\protect\citeauthoryear{{Elmegreen} \& {Scalo}}{{Elmegreen} \&
  {Scalo}}{2004}]{elmegreen2004}
{Elmegreen} B.~G.,  {Scalo} J.,  2004, \mn@doi [Annual Review of Astronomy and
  Astrophysics] {10.1146/annurev.astro.41.011802.094859}, \href
  {https://ui.adsabs.harvard.edu/\#abs/2004ARA&A..42..211E} {42, 211}

\bibitem[\protect\citeauthoryear{{Federrath} \& {Banerjee}}{{Federrath} \&
  {Banerjee}}{2015}]{Federrath2015}
{Federrath} C.,  {Banerjee} S.,  2015, \mn@doi [\mnras] {10.1093/mnras/stv180},
  \href {http://adsabs.harvard.edu/abs/2015MNRAS.448.3297F} {448, 3297}

\bibitem[\protect\citeauthoryear{{Federrath} \& {Klessen}}{{Federrath} \&
  {Klessen}}{2012}]{federrath2012}
{Federrath} C.,  {Klessen} R.~S.,  2012, \mn@doi [\apj]
  {10.1088/0004-637X/761/2/156}, \href
  {http://adsabs.harvard.edu/abs/2012ApJ...761..156F} {761, 156}

\bibitem[\protect\citeauthoryear{{Federrath}, {Klessen}  \&
  {Schmidt}}{{Federrath} et~al.}{2008}]{Federrath08}
{Federrath} C.,  {Klessen} R.~S.,   {Schmidt} W.,  2008, \apjl, 688, L79

\bibitem[\protect\citeauthoryear{{Federrath}, {Roman-Duval}, {Klessen},
  {Schmidt}  \& {Mac Low}}{{Federrath} et~al.}{2010}]{Federrath10}
{Federrath} C.,  {Roman-Duval} J.,  {Klessen} R.~S.,  {Schmidt} W.,   {Mac Low}
  M.-M.,  2010, \aap, 512, A81

\bibitem[\protect\citeauthoryear{{Fryer}, {Ellinger}, {Young}  \&
  {Vance}}{{Fryer} et~al.}{2017}]{fryer2017}
{Fryer} C.~L.,  {Ellinger} C.,  {Young} P.~A.,   {Vance} G.,  2017, in
  {Marcowith} A.,  {Renaud} M.,  {Dubner} G.,  {Ray} A.,   {Bykov} A.,  eds,
  IAU Symposium Vol. 331, Supernova 1987A:30 years later - Cosmic Rays and
  Nuclei from Supernovae and their Aftermaths. pp 86--95,
  \mn@doi{10.1017/S174392131700641X}

\bibitem[\protect\citeauthoryear{{Hartigan} et~al.,}{{Hartigan}
  et~al.}{2016}]{2016ApJ...823..148H}
{Hartigan} P.,  et~al., 2016, \mn@doi [\apj] {10.3847/0004-637X/823/2/148},
  \href {http://adsabs.harvard.edu/abs/2016ApJ...823..148H} {823, 148}

\bibitem[\protect\citeauthoryear{{Inoue}, {Yamazaki}  \& {Inutsuka}}{{Inoue}
  et~al.}{2009}]{inoue2009}
{Inoue} T.,  {Yamazaki} R.,   {Inutsuka} S.-i.,  2009, \mn@doi [\apj]
  {10.1088/0004-637X/695/2/825}, \href
  {https://ui.adsabs.harvard.edu/\#abs/2009ApJ...695..825I} {695, 825}

\bibitem[\protect\citeauthoryear{{Kolmogorov}}{{Kolmogorov}}{1941}]{kolmogorov1941}
{Kolmogorov} A.,  1941, Akademiia Nauk SSSR Doklady, \href
  {http://adsabs.harvard.edu/abs/1941DoSSR..30..301K} {30, 301}

\bibitem[\protect\citeauthoryear{{Kritsuk}, {Norman}, {Padoan}  \&
  {Wagner}}{{Kritsuk} et~al.}{2007}]{kritsuk2007}
{Kritsuk} A.~G.,  {Norman} M.~L.,  {Padoan} P.,   {Wagner} R.,  2007, \mn@doi
  [\apj] {10.1086/519443}, \href
  {http://adsabs.harvard.edu/abs/2007ApJ...665..416K} {665, 416}

\bibitem[\protect\citeauthoryear{{Kritsuk} et~al.,}{{Kritsuk}
  et~al.}{2011}]{kritsuk2011b}
{Kritsuk} A.~G.,  et~al., 2011, \mn@doi [\apj] {10.1088/0004-637X/737/1/13},
  \href {https://ui.adsabs.harvard.edu/\#abs/2011ApJ...737...13K} {737, 13}

\bibitem[\protect\citeauthoryear{{Kritsuk}, {Ustyugov}  \& {Norman}}{{Kritsuk}
  et~al.}{2017}]{kritsuk2017}
{Kritsuk} A.~G.,  {Ustyugov} S.~D.,   {Norman} M.~L.,  2017, \mn@doi [New
  Journal of Physics] {10.1088/1367-2630/aa7156}, \href
  {https://ui.adsabs.harvard.edu/\#abs/2017NJPh...19f5003K} {19, 065003}

\bibitem[\protect\citeauthoryear{{Lebedev} et~al.,}{{Lebedev}
  et~al.}{2002}]{2002ApJ...564..113L}
{Lebedev} S.~V.,  et~al., 2002, \mn@doi [\apj] {10.1086/324183}, \href
  {http://adsabs.harvard.edu/abs/2002ApJ...564..113L} {564, 113}

\bibitem[\protect\citeauthoryear{{Lebedev} et~al.,}{{Lebedev}
  et~al.}{2014}]{Lebedev2014}
{Lebedev} S.~V.,  et~al., 2014, \mn@doi [Physics of Plasmas]
  {10.1063/1.4874334}, \href
  {http://adsabs.harvard.edu/abs/2014PhPl...21e6305L} {21, 056305}

\bibitem[\protect\citeauthoryear{{Lebedev}, {Frank}  \& {Ryutov}}{{Lebedev}
  et~al.}{ress}]{Lebedev2019}
{Lebedev} S.~V.,  {Frank} A.,   {Ryutov} D.~D.,  \noop{3001}in press, Reviews
  of Modern Physics

\bibitem[\protect\citeauthoryear{{Mac Low} \& {Klessen}}{{Mac Low} \&
  {Klessen}}{2004}]{maclow2004}
{Mac Low} M.-M.,  {Klessen} R.~S.,  2004, \mn@doi [Reviews of Modern Physics]
  {10.1103/RevModPhys.76.125}, \href
  {http://adsabs.harvard.edu/abs/2004RvMP...76..125M} {76, 125}

\bibitem[\protect\citeauthoryear{{Meinecke} et~al.,}{{Meinecke}
  et~al.}{2014}]{2014NatPh..10..520M}
{Meinecke} J.,  et~al., 2014, \mn@doi [Nature Physics] {10.1038/nphys2978},
  \href {http://adsabs.harvard.edu/abs/2014NatPh..10..520M} {10, 520}

\bibitem[\protect\citeauthoryear{{Meinecke} et~al.,}{{Meinecke}
  et~al.}{2015}]{Meinecke2015}
{Meinecke} J.,  et~al., 2015, Proceedings of the National Academy of Science,
  112, 8211

\bibitem[\protect\citeauthoryear{{Miyoshi} \& {Kusano}}{{Miyoshi} \&
  {Kusano}}{2005}]{miyoshi2005}
{Miyoshi} T.,  {Kusano} K.,  2005, \mn@doi [Journal of Computational Physics]
  {10.1016/j.jcp.2005.02.017}, \href
  {http://adsabs.harvard.edu/abs/2005JCoPh.208..315M} {208, 315}

\bibitem[\protect\citeauthoryear{{Molina}, {Glover}, {Federrath}  \&
  {Klessen}}{{Molina} et~al.}{2012}]{molina2012}
{Molina} F.~Z.,  {Glover} S.~C.~O.,  {Federrath} C.,   {Klessen} R.~S.,  2012,
  \mn@doi [\mnras] {10.1111/j.1365-2966.2012.21075.x}, \href
  {http://adsabs.harvard.edu/abs/2012MNRAS.423.2680M} {423, 2680}

\bibitem[\protect\citeauthoryear{{Offner} \& {Liu}}{{Offner} \&
  {Liu}}{2018}]{offner2018}
{Offner} S.~S.~R.,  {Liu} Y.,  2018, \mn@doi [Nature Astronomy]
  {10.1038/s41550-018-0566-1}, \href
  {http://adsabs.harvard.edu/abs/2018NatAs...2..896O} {2, 896}

\bibitem[\protect\citeauthoryear{{Padoan}, {Nordlund}  \& {Jones}}{{Padoan}
  et~al.}{1997}]{padoan1997}
{Padoan} P.,  {Nordlund} A.,   {Jones} B.~J.~T.,  1997, \mn@doi [\mnras]
  {10.1093/mnras/288.1.145}, \href
  {http://adsabs.harvard.edu/abs/1997MNRAS.288..145P} {288, 145}

\bibitem[\protect\citeauthoryear{{Padoan}, {Jimenez}, {Nordlund}  \&
  {Boldyrev}}{{Padoan} et~al.}{2004}]{padoan2004}
{Padoan} P.,  {Jimenez} R.,  {Nordlund} {\AA}.,   {Boldyrev} S.,  2004, \mn@doi
  [Physical Review Letters] {10.1103/PhysRevLett.92.191102}, \href
  {http://adsabs.harvard.edu/abs/2004PhRvL..92s1102P} {92, 191102}

\bibitem[\protect\citeauthoryear{{Passot} \& {V{\'a}zquez-Semadeni}}{{Passot}
  \& {V{\'a}zquez-Semadeni}}{1998}]{passot1998}
{Passot} T.,  {V{\'a}zquez-Semadeni} E.,  1998, \mn@doi [\pre]
  {10.1103/PhysRevE.58.4501}, \href
  {http://adsabs.harvard.edu/abs/1998PhRvE..58.4501P} {58, 4501}

\bibitem[\protect\citeauthoryear{{Radice}, {Abdikamalov}, {Ott}, {M{\"o}sta},
  {Couch}  \& {Roberts}}{{Radice} et~al.}{2018}]{radice2018}
{Radice} D.,  {Abdikamalov} E.,  {Ott} C.~D.,  {M{\"o}sta} P.,  {Couch} S.~M.,
   {Roberts} L.~F.,  2018, \mn@doi [Journal of Physics G Nuclear Physics]
  {10.1088/1361-6471/aab872}, \href
  {https://ui.adsabs.harvard.edu/\#abs/2018JPhG...45e3003R} {45, 053003}

\bibitem[\protect\citeauthoryear{{Remington}, {Drake}  \& {Ryutov}}{{Remington}
  et~al.}{2006}]{2006RvMP...78..755R}
{Remington} B.~A.,  {Drake} R.~P.,   {Ryutov} D.~D.,  2006, \mn@doi [Reviews of
  Modern Physics] {10.1103/RevModPhys.78.755}, \href
  {http://adsabs.harvard.edu/abs/2006RvMP...78..755R} {78, 755}

\bibitem[\protect\citeauthoryear{{Roy}, {Bharadwaj}, {Dutta}  \&
  {Chengalur}}{{Roy} et~al.}{2009}]{roy2009}
{Roy} N.,  {Bharadwaj} S.,  {Dutta} P.,   {Chengalur} J.~N.,  2009, \mn@doi
  [\mnras] {10.1111/j.1745-3933.2008.00591.x}, \href
  {https://ui.adsabs.harvard.edu/\#abs/2009MNRAS.393L..26R} {393, L26}

\bibitem[\protect\citeauthoryear{{Scalo}, {V{\'a}zquez-Semadeni}, {Chappell}
  \& {Passot}}{{Scalo} et~al.}{1998}]{scalo1998}
{Scalo} J.,  {V{\'a}zquez-Semadeni} E.,  {Chappell} D.,   {Passot} T.,  1998,
  \mn@doi [\apj] {10.1086/306099}, \href
  {http://adsabs.harvard.edu/abs/1998ApJ...504..835S} {504, 835}

\bibitem[\protect\citeauthoryear{{Schmidt}, {Federrath}, {Hupp}, {Kern}  \&
  {Niemeyer}}{{Schmidt} et~al.}{2009}]{schmidt2009}
{Schmidt} W.,  {Federrath} C.,  {Hupp} M.,  {Kern} S.,   {Niemeyer} J.~C.,
  2009, \mn@doi [\aap] {10.1051/0004-6361:200809967}, \href
  {http://adsabs.harvard.edu/abs/2009A%26A...494..127S} {494, 127}

\bibitem[\protect\citeauthoryear{{Shu} \& {Osher}}{{Shu} \&
  {Osher}}{1988}]{shu1988}
{Shu} C.-W.,  {Osher} S.,  1988, \mn@doi [Journal of Computational Physics]
  {10.1016/0021-9991(88)90177-5}, \href
  {http://adsabs.harvard.edu/abs/1988JCoPh..77..439S} {77, 439}

\bibitem[\protect\citeauthoryear{{Suzuki-Vidal} et~al.,}{{Suzuki-Vidal}
  et~al.}{2015}]{2015ApJ...815...96S}
{Suzuki-Vidal} F.,  et~al., 2015, \mn@doi [\apj] {10.1088/0004-637X/815/2/96},
  \href {http://adsabs.harvard.edu/abs/2015ApJ...815...96S} {815, 96}

\bibitem[\protect\citeauthoryear{{Toro}, {Spruce}  \& {Speares}}{{Toro}
  et~al.}{1994}]{toro1994}
{Toro} E.~F.,  {Spruce} M.,   {Speares} W.,  1994, \mn@doi [Shock Waves]
  {10.1007/BF01414629}, \href
  {http://adsabs.harvard.edu/abs/1994ShWav...4...25T} {4, 25}

\bibitem[\protect\citeauthoryear{{Usmanov}, {Goldstein}  \&
  {Matthaeus}}{{Usmanov} et~al.}{2014}]{Usmanov2014}
{Usmanov} A.~V.,  {Goldstein} M.~L.,   {Matthaeus} W.~H.,  2014, \mn@doi [\apj]
  {10.1088/0004-637X/788/1/43}, \href
  {http://adsabs.harvard.edu/abs/2014ApJ...788...43U} {788, 43}

\bibitem[\protect\citeauthoryear{{Vazquez-Semadeni}}{{Vazquez-Semadeni}}{1994}]{Vazquez-Semadeni1994}
{Vazquez-Semadeni} E.,  1994, \mn@doi [\apj] {10.1086/173847}, \href
  {http://adsabs.harvard.edu/abs/1994ApJ...423..681V} {423, 681}

\bibitem[\protect\citeauthoryear{{Vishniac}}{{Vishniac}}{1994}]{vishniac1994}
{Vishniac} E.~T.,  1994, \mn@doi [\apj] {10.1086/174231}, \href
  {http://adsabs.harvard.edu/abs/1994ApJ...428..186V} {428, 186}

\makeatother
\end{thebibliography}

% Don't change these lines
\bsp	% typesetting comment
\label{lastpage}
\end{document}